\documentclass[11pt]{article}

\usepackage{xcolor}\usepackage[letterpaper,top=3.5cm,bottom=3cm,left=3cm,right=3cm,marginparwidth=1.75cm]{geometry}

\usepackage{amsmath}
\usepackage{graphicx}
\usepackage[colorlinks=true, allcolors=black]{hyperref}

\title{\LARGE{\textbf{Characterization and modeling of spiking and bursting in experimental $\textrm{NbO}_\textrm{x}$ neuron}}}

\author{\small{Marie Drouhin$^{1,2}$, Shuai Li$^{1}$, Matthieu Grelier$^{1}$, Sophie Collin$^{1}$, Florian Godel$^{1}$, Robert G. Elliman$^{3}$, }\\
\small{ Bruno Dlubak$^{1}$, Juan Trastoy$^{1}$,  Damien Querlioz$^{2}$, Julie Grollier$^{1}$}}

\date{\scriptsize{$^1$: Unité Mixte de Physique CNRS, Thales, Université Paris-Saclay, 91767 Palaiseau, France\\
$^2$: Université Paris-Saclay, CNRS, Centre de Nanosciences et de Nanotechnologies, 91120 Palaiseau, France.\\
$^{3}$: Department of Electronic Material Engineering, Research School of Physics, The Australian National University, Canberra, ACT 2601, Australia}}

\begin{document}
\maketitle
\begin{abstract}
Hardware spiking neural networks hold the promise of realizing artificial intelligence with high energy efficiency. In this context,  solid-state and scalable memristors can be used to mimic biological neuron characteristics.  However, these devices show limited neuronal behaviors and have to be integrated in more complex circuits to implement the rich dynamics of biological neurons. Here we studied a $\textrm{NbO}_\textrm{x}$ memristor neuron that is capable of emulating numerous neuronal dynamics, including tonic spiking, stochastic spiking, leaky-integrate-and-fire features, spike latency, temporal integration. The device also exhibits phasic bursting, a property that has scarcely been observed and studied in solid-state nano-neurons.   We show that we can reproduce and understand this particular response through simulations using non-linear dynamics.  These results show that a single $\textrm{NbO}_\textrm{x}$ device is sufficient to emulate a collection of rich neuronal dynamics that paves a path forward for realizing scalable and energy-efficient neuromorphic computing paradigms. 
\end{abstract}

\section{Introduction}
As the interest in Artificial Intelligence (AI) grows,  spiking neural networks offer an energy-efficient, hardware-compatible, and event-driven alternative to  conventional artificial neural networks
\cite{pfeiffer2018deep}, particularly adapted for processing sensory and dynamical data. 
Hardware spiking neurons can be realized solely using complementary metal oxide semiconductor (CMOS) technology, but this type of implementation suffers from a lack of scalability \cite{thakur2018large}. This limitation explains the growing interest in the realization of new devices that feature neuronal behavior and that can be scaled easily \cite{kendall2020building, zidan2018future}. However, researchers face  the choice between single, scalable nanodevices  that exhibit  a limited range of neuronal responses and more complex neurons  that offer more diverse behavior but limited scalability. Having more diverse behavior provides the potential of reproducing the brain's computational power to its full extent. 
Biological neurons may indeed exhibit different types of spiking responses, as well as bursting responses, where a neuron produces multiple spikes in response to an input pulse. 
A neuron implementing a highly simplified response will fail to provide the complexity required to emulate neurobiology. 
For example, the bursting response is believed to be of importance for ensuring reliable communication and synchronization between neurons \cite{izhikevich2007dynamical} \cite{payeur2021burst}.
Therefore, considerable effort has been devoted to realizing new scalable devices with diverse neuronal characteristics \cite{izhikevich2007dynamical} \cite{yi2018biological} \cite{kumar2020third}. 

A leading idea to 
engineer this new type of devices is to exploit the intrinsic physics of nanoscale materials to implement neurons \cite{markovic2020physics,ielmini2021brain, wang2020resistive, lee2020nanoscale, xi2020memory}. A large number of devices have been studied for their neuronal applications  \cite{choi2020emerging} \cite{yang2019memristive} \cite{yang2021nonlinearity}: phase change neuron \cite{tuma2016stochastic}, valence change neuron \cite {woo2017dual} \cite{wang2018handwritten},   electrochemical metallization neuron \cite{zhang2017artificial}, diffusive neuron \cite{wang2018fully}, Mott insulator neuron \cite{stoliar2017leaky}, and spintronic neuron \cite{grollier2020neuromorphic}. 
Within these examples, metal/insulator/metal structures based on transition metal oxides such as $VO_x$ and $NbO_x$ are particularly promising candidates, as they exhibit reliable threshold switching and current-controlled negative differential resistance (NDR) characteristics. $NbO_x$ memristor neurons feature high endurance \cite{li2015high} and have been shown to be capable of leaky integrate-and-fire,  all-or-nothing spiking and chaotic oscillations \cite{kumar2017chaotic}. This type of device has also been  used to implement dynamic, logic, and multiplicative gain modulation \cite{duan2020spiking}. However, the behavior of a single device is nowhere near as complex as a real biological neuron. To obtain more sophisticated behavior, complex devices featuring  multiple electrophysical processes have to be created \cite{kumar2020third}, which can be challenging to model and control precisely.
Alternatively, several neuronal devices can be used together in appropriately engineered circuits \cite{yi2018biological}.

In this work, we fabricate and characterize memristor neurons based on a simple Pt/$\textrm{Nb}_\textrm{2}\textrm{O}_\textrm{5}$/ Ti/Pt stack with current inputs and output voltage shapes that are close to the shape of a biological action potential, thanks to the effect of an inductance. These devices are straightforward to model with physics equations, and simultaneously,  feature  multiple computational properties such as tonic spiking, stochastic spiking, spike latency, leaky-and-fire integration (LIF), all-or-nothing firing, and phasic bursting. These neuron-like dynamics can be modelled and understood through physical equations and standard non-linear dynamics.\\

\section{Fabrication and method}

 $NbO_x$ memristors, comprising 5 $\mu m \times 5  \mu m$ cross-point structures,  were fabricated by successive film deposition and patterning. A 4-nm Ti adhesion layer  and a 25-nm thick Pt layer were first deposited on a SiO$_2$/Si substrate by  electron-beam evaporation. These layers were subsequently patterned using optical lithography and ion-beam etching to define the bottom electrodes.    A 30 nm  $Nb_2O_5$ layer was then  deposited onto the bottom electrodes using radio-frequency sputtering from a $Nb_2O_5$ target at room temperature in an Ar ambient. The metal-oxide-metal device was completed by adding a top electrode (10 nm Ti - 25 nm Pt) deposited by electron beam evaporation.  

For electrical measurements, the bottom electrode was connected to  ground and the source applied to the top electrode. I-V characteristics were measured with a Keysight B1500A Semiconductor Device Analyzer after   current-controlled electroforming with a positive polarity. Pulse measurements were performed using an Agilent 81160A pulse generator and a voltage-pulse to current-pulse converter (see Supplementary Materials figure~\ref{voltage-current_converter} for the exact structure). The spiking behavior was monitored on a 2~GHz-bandwidth Keysight MSOS204A oscilloscope. All measurements were performed with a DC probe station. 

Before the electroforming process, the resistance of the device was about 4 M$\Omega$ at 0.3 V. Electroforming was achieved by the application of a current ramp from 0 to 0.5~mA to the device (see Supplementary Materials figure~\ref{electroforming}). After this step, the device resistance was reduced to 93 k$\Omega$ at 0.3 V.

The simulations presented in figure~\ref{Fig1}  were computed with LTSpice using the electrical circuit presented in figure~\ref{Fig1}a based on  the Newton law of cooling and the Poole-Frenkel effect (see equations \eqref{eq: R_d} and \eqref{Eq: Newton's Cooling Law} below), with a 5~ns time step.  The values of all parameters used in these simulations are listed in table~\hyperlink{table}{1}. The temperature evolution was implemented in LTSpice following guidelines described in the supporting information of \cite{li2019origin}.  
The simulations shown in figure~\ref{Fig4} were executed in Python with a Runge-Kutta solver of order 5  and a timestep of 50 ps using equations \eqref{eq: R_d} and \eqref{Eq: Newton's Cooling Law} described below.

\section{Results}

\begin{figure} [h]
\centering
\includegraphics[width=1\textwidth]{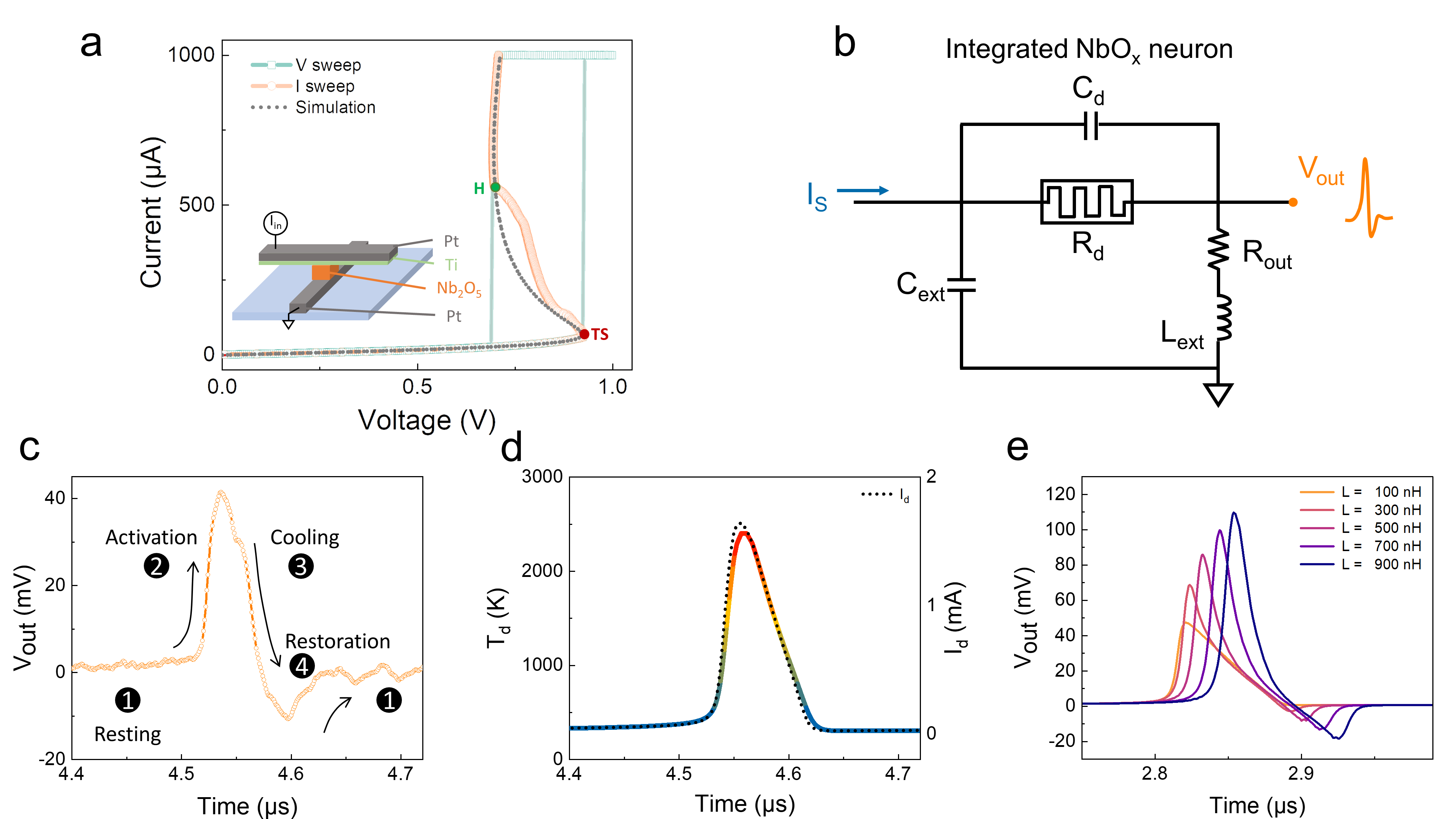}
\caption{\label{Fig1} a. Measured (square symbols) and simulated (dots) I-V characteristics. The V sweep and I sweep correspond respectively to the voltage-controlled and current-controlled I-V characteristics. The hold point H is indicated in green and the threshold switching point TS in red. The inset shows a sketch of the structure of the device. b. Circuit diagram of the integrated $NbO_x$ spiking neuron where $C_{ext}$ and $L_{ext}$ are respectively a parasitic capacitance and inductance. c.  Measurement of a single spike of a $NbO_x$ neuron, with the four stages of an action potential indicated.  d. Simulated spiking dynamics of the $NbO_x$ neuron temperature $T_d$ (colored curve) and current $I_d$ (dots) for a constant input current of $180~\mu A$. e. Simulation of the output voltage shape with respect to the value of the circuit inductance for a constant input current of $180~\mu A$.}
\end{figure}

The quasistatic I-V characteristics of our device are shown in figure~\ref{Fig1}a, highlighting the current-controlled S-shaped Negative Differential Resistance (NDR) response, characteristic of  a voltage-controlled Threshold Switching (TS). 
Two characteristic values are included on the graph. The first one is the Threshold Switching point (called TS in figure~\ref{Fig1}a), where the slope of the current-controlled I-V characteristic goes from positive to negative. This point also coincides with the abrupt transition from a high-resistance state to a low-resistance state under voltage controlled transitions. The second  is the hold point H, where the differential resistance becomes positive again.  \\

The physical basis of this behavior has been under debate but is generally understood to arise from an increase in the oxide electrical conductivity due to local Joule heating.  Indeed, Gibson \cite{gibson2018designing} has shown that the NDR response can arise from any mechanism that gives rise to a superlinear increase in conductivity with temperature.  In the case of $NbO_x$,  some authors initially attributed it to a characteristic   insulator-to-metal  transition (IMT) in $NbO_2$ \cite{pickett2012sub}, but it is now generally accepted that it arises from a  trap-assisted transport mechanism, such as Poole-Frenkel conduction 
\cite{slesazeck2015physical, wang2018transient}. 

In the case
of the Poole-Frenkel effect, a filament of oxygen vacancies connects both electrodes after electroforming. The oxygen vacancies act as potential traps for electrons.
If an electric field is applied to the device, the energy profile of the conduction band in the oxide around the traps becomes asymmetric. Trapped electrons are then able to be thermally injected  into the conduction band, leading to the traditional Poole-Frenkel equation for the device resistance $R_d$ as a function of the temperature $T_d$ and the voltage $V_d$ across the device:   
\begin{equation}
    R_d = R_0 exp \left(\frac{E_a -q\sqrt{\frac{q V_d}{\pi \epsilon_0 \epsilon_r d}}}{k_{B} T_d} \right),
\label{eq: R_d}
\end{equation}
where $E_a$ is the activation energy associated with the carrier trap level, $\epsilon_0$  the vacuum permittivity,   $\epsilon_r$ the relative permittivity of $NbO_x$, $q$  the elementary charge, and $d$  the thickness of the oxide film. $V_d$ is the device voltage and $T_d$ is the temperature of the active device volume \cite{slesazeck2015physical}.
The occurrence of electrical current through the filament results in a positive feedback, where Joule heating raises the local temperature $T_d$, reducing the device resistance further \cite{poole1916viii, frenkel1938pre}.
This phenomenon can be modeled from a lumped element model of the device, where the Newton's Cooling Law is used to describe the evolution of the temperature,
\begin{equation}
    \frac{d T_d}{dt} = \frac{V_d^2}{R_d C_{th}} - \frac{T_d - T_{amb}}{C_{th} R_{th}} 
\label{Eq: Newton's Cooling Law}
\end{equation}
where $T_{amb}$ is the room temperature, and $C_{th}$ and $R_{th}$ are respectively the thermal capacitor and resistor.
We simulated the I-V curve of our device using these equations (see methods). The simulation results presented 
 with a dotted line in figure~\ref{Fig1}a show that the model reproduces the experimental data.

Figure~\ref{Fig1}b presents the simple experimental setup used to measure the spiking behavior of  neurons. In this circuit, $R_d$ is the device resistance described by equation~\ref{eq: R_d} and $C_d$ is the intrinsic device capacitance arising from its metal/insulator/metal structure. $C_{ext}$ and $L_{ext}$ respectively account for parasitic capacitance and inductance of the measurement set-up. $R_{out}$ is an external resistor of 25 Ohms across which the output voltage is measured.  The input of the circuit is a current, and the output is a voltage, in line with the biological configuration.
Figure~\ref{Fig1}c shows an experimentally measured spike, observed by applying a constant $150~\mu A$ current input to the circuit. The shape of the output spike strongly resembles that of a biological neuron, with an initial  depolarization followed by hyperpolarization: starting from a resting phase, the output voltage increases rapidly during the activation phase, and then decreases to become negative before rising again to the resting phase.  

To understand this behavior, figure~\ref{Fig1}d shows simulations of the current $I_d$ flowing through the device (dots) and the simulated temperature $T_d$ of the active device volume (colored curve) during a spike, using the LTSpice model of our experiment and a current input of $180~\mu A$ (see methods). These responses are clearly correlated, with both curves exhibiting a  rapid increase and  a slower decrease, which can be explained as follows. The device is initially in an insulating state. When a constant  current is applied, 
the capacitance $C_{ext}$ charges and the voltage across the device increases until it approaches the threshold voltage, at which point the device resistance drops, producing the increase in current and temperature evident in figure \ref{Fig1}d.  This discharges the capacitor, reducing the device voltage to the point where the memristor reverts to its subthreshold resistance.  The transition to a high resistance state causes a reduction in current and temperature, ending the spike response.  

The restoration part of the neuron-like voltage spike is seen in the output voltage but not in the current and temperature curves; this is due to the presence of a parasitic inductance (see figure~\ref{Fig1}b). The device intrinsic capacitance $C_d$ is  small, and the current in that branch is also  small. Therefore, the current going through the inductance $L_{ext}$ and the output resistor $R_{out}$ (figure~\ref{Fig1}d), is close to that going through the neuron $R_d$. Because the voltage across the inductance opposes the variations of the current, it is first positive and then negative. The output voltage is  the sum of two terms,  $V_{out} = R_{out} i_{out} + L_{ext} \frac {d i_{out}}{dt}$: if the inductance is large enough, the output voltage is first positive (during the activation  part) and then  decreases until it becomes negative (during the cooling and restoration parts). This mechanism explains the results shown in figure~\ref{Fig1}e, where the evolution of the shape of the pulse with respect to the circuit inductance is simulated. When the inductance is smaller than $100~n H$, it has a negligible impact on the output voltage (showed in figure~\ref{Fig1}c); for higher inductance values, a restoration phase is observed. \\

\begin{figure} [h]
\centering
\includegraphics[width=1\textwidth]{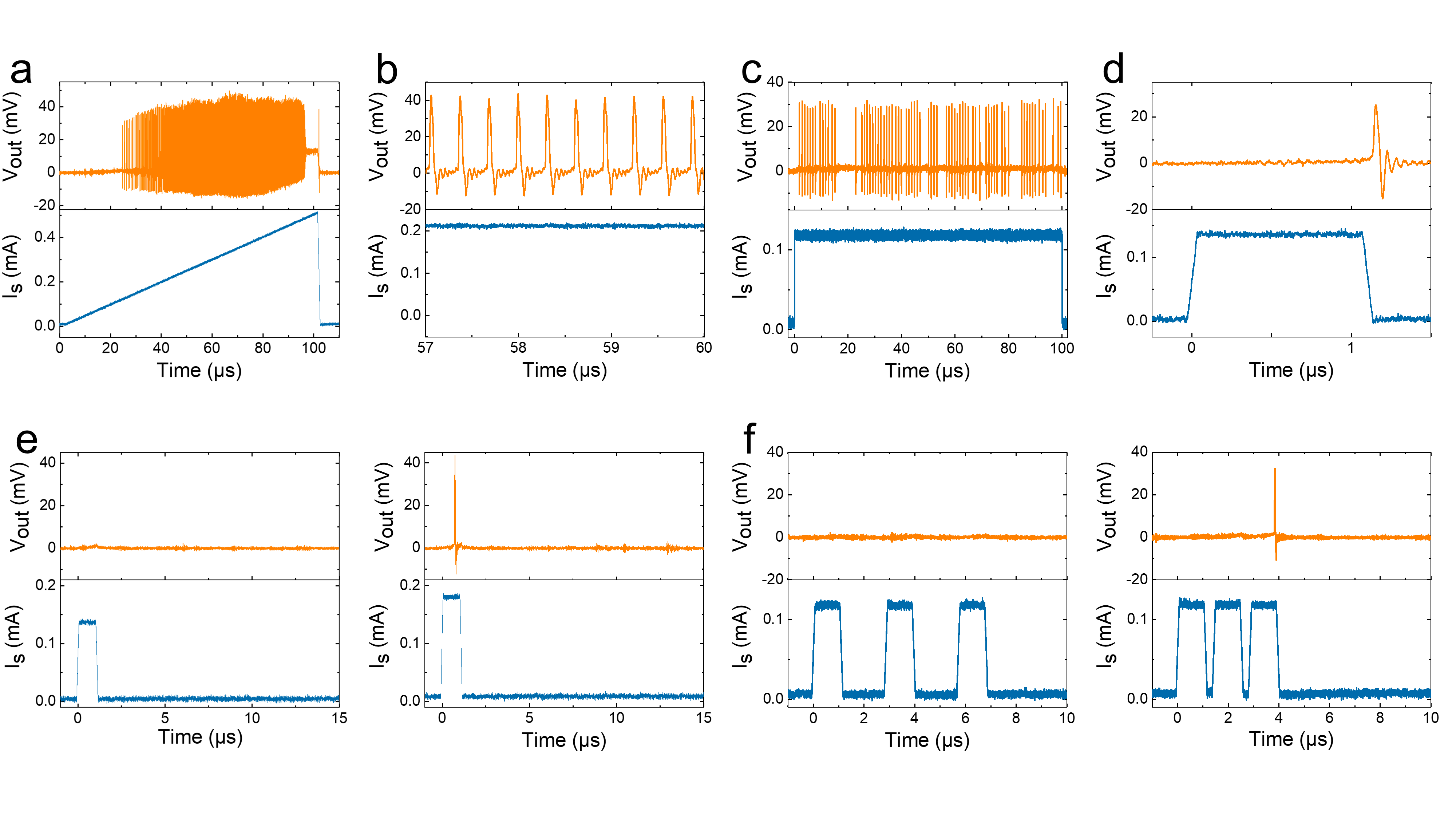}
\caption{\label{Fig2} a: $NbO_x$ neuron output as a function of input current amplitude. A 99 $\mu s$ current ramp from 0 to 0.46 mA and 1 $\mu s$ fall time is applied to the device. b: Tonic spiking. The neuron receives a constant input current of 0.2 mA. c: Stochastic spiking obtained with a current of 0.109 mA.  d: Spike latency. A pulse with a duration of 1 $\mu s$, a rise time and fall time of both 100 ns and an amplitude of 0.131 mA  is applied to the neuron. e: Spatial integration. Comparison between two figures where a pulse of duration of 1 $\mu s$ with a rise time and fall time of both 100 ns are applied to the neuron. The input current value is 0.13 mA on the left and 0.17 mA on the right. f: Temporal integration. Three pulses of of duration of 1 $\mu s$ with a rise time and fall time of both 100 ns and of amplitude 0.110 mA are applied to the neuron. The frequency is 0.35 MHz on the left and 0.7 MHz on the right.}
\end{figure}

Having analyzed the $NbO_x$ neuron spike shape, we now explore its computational properties.  
Figure~\ref{Fig2}a shows the neuron behavior when a current ramp is applied at its input. For low currents the neuron does not spike, as the NDR behavior needed for spike generation does not appear until the current reaches the TS point in figure~\ref{Fig1}a.  Above this threshold current, the neuron spikes with increasing frequency until the current exceeds the hold value (H) of figure~\ref{Fig1}a, above which the NDR disappears as well as the related spiking behavior. This characteristic is reproduced in simulations in Supplementary Materials figure~\ref{ramp}. 

When the input is constant and lies between the threshold current and the hold current, the neuron spikes with a constant frequency, a behavior called tonic spiking for biological neurons, as shown  in figure~\ref{Fig2}b (and reproduced in simulations in Supplementary Materials figure~\ref{tonic}).
Close to the threshold current, the behavior is stochastic, as shown in figure~\ref{Fig2}c, as can be expected from a thermally-driven process, but with a non-random occurrence of spiking events, that can be described by quiet periods followed by bursts of spikes with constant frequency. Due to input current noise, the neuron output indeed fluctuates between its below-threshold behavior (no spikes) and its above-threshold behaviors (spikes with a constant frequency). This stochastic bursting behavior is reminiscent of biological neuron bursting and could be exploited for computations and learning in hardware circuits \cite{payeur2021burst}.

The neuron also exhibits spike latency, as evidenced in figure~\ref{Fig2}d for a $1~\mu s$-duration pulse  applied to the device. During the whole duration of the input, the output voltage does not show any significant response. However, once the pulse is  back to zero, the neuron spikes. 
This effect can be explained naturally within the  context of the above model. Indeed, when the current pulse is applied long enough for the temperature to activate the Poole-Frenkel effect, the positive feedback mechanism starts and the temperature keeps increasing even as the source stops, giving rise to  spike latency. This behavior is simulated in Supplementary Materials figure~\ref{latency}.

Moreover, the neuron may exhibit all-or-nothing behavior. In figure~\ref{Fig2}e, two pulses with the same duration of $1~\mu s$ are applied to the neuron with different current input values: 0.13~mA for the left figure  and 0.17~mA for the right one. The first pulse is not sufficient to make the neuron spike, but a slight variation of the output voltage can be observed. The second pulse is high enough to make the neuron spike, as the value of the current has been increased. In the context of a spiking neural network, this all-or-nothing behavior allows triggering a neuron only when a sufficient number of spikes (with below-threshold amplitude) arrives simultaneously at its input, thus filtering meaningful signal only, a behavior akin to spatial summation. This behavior is reproduced with simulations in figure~\ref{all-or-nothing}.

Finally, figure~\ref{Fig2}f displays a different situation where three pulses of identical duration ($1 \mu s$) and peak current  (0.11~mA) are applied. On the left, the input frequency of 0.35~MHz is not high enough for the neuron to spike, contrary to the right panel in which the frequency is increased to 0.7 MHz, allowing it to spike. This behavior indicates a  frequency-dependent temporal summation by the neuron, reproduced with simulations in Supplementary Materials figure~\ref{LIF}. This typical leaky-integrate-and-fire behavior is particularly adapted  for spiking neural networks where frequency encodes the information.\\

\begin{figure} [h]
\centering
\includegraphics[width=1.1\textwidth]{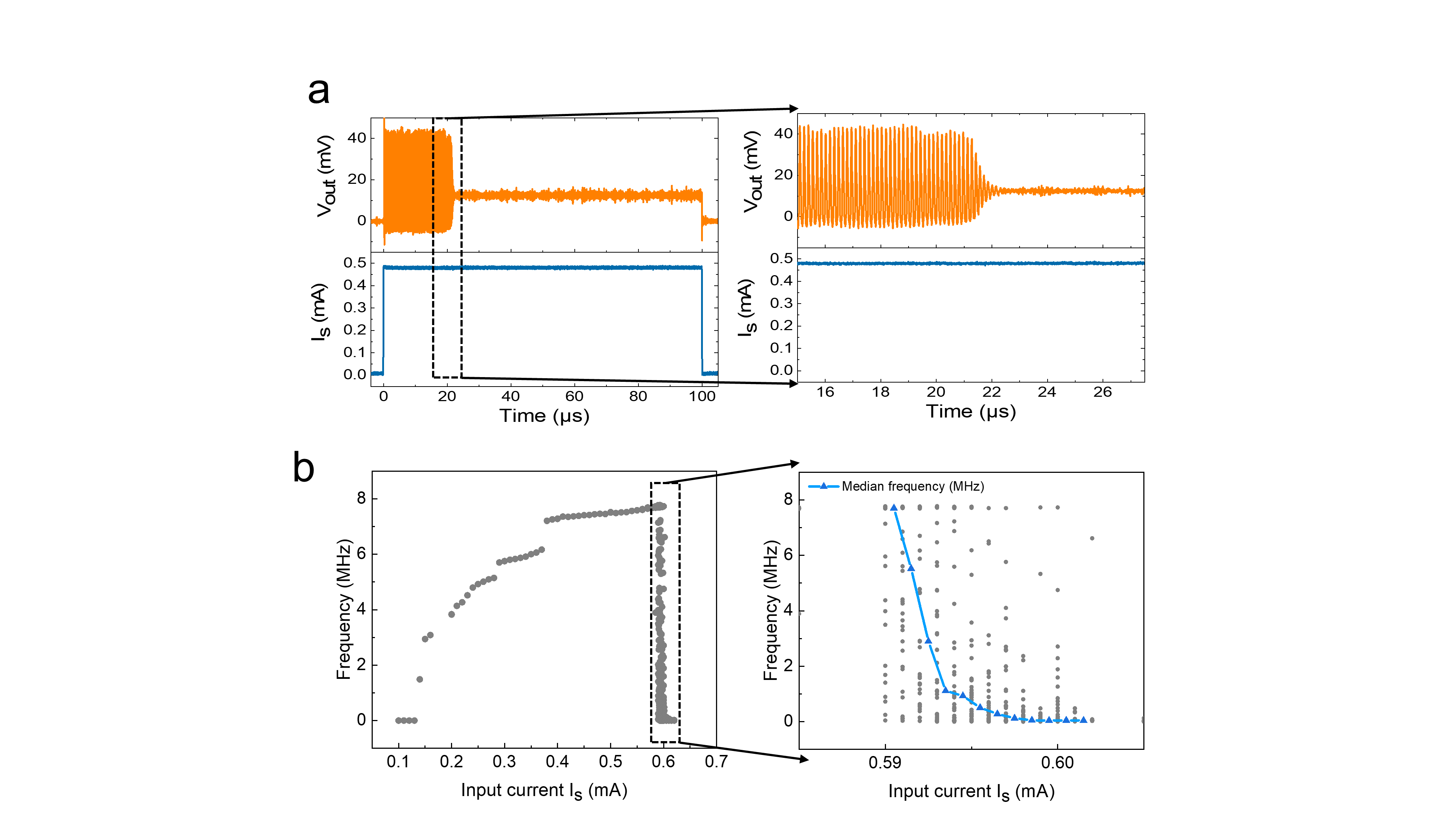}
\caption{\label{Fig3} a: Example of phasic bursting of the output voltage as a function of time. A current input of amplitude 0.47 mA is applied. The right panel zooms on the end of the phasic bursting.  b: Left: Variation of the average frequency as a function of the input current. Right: Zoom on the phasic bursting regime, in order to get a statistical understanding of the phenomenon. In blue, the median frequency computed from the different average frequencies (grey dots) is plotted.}
\end{figure}

While most of the spiking features presented in figure~\ref{Fig2} have been reported for various types of solid-state neurons \cite{kumar2020third, woo2017dual, zhang2017artificial,stoliar2017leaky}, figure~\ref{Fig3} shows that our simple $NbO_x$ neuron exhibits a behavior observed in biological neurons and scarcely investigated in memristive systems, named phasic bursting \cite{kumar2020third}.  In this case, for a constant input current just above the hold point (see figure~\ref{Fig1}a), the neuron starts to spike before stopping abruptly, as shown in figure~\ref{Fig3}a. This situation differs from figure~\ref{Fig2}a, where a current ramp was applied. In figure~\ref{Fig2}a, the neuron stopped spiking near the end of the input ramp, because the input current ended well above the Hold current (H point in figure~\ref{Fig1}). In figure~\ref{Fig3}a, the input is now constant and the neuron still spikes before stopping abruptly.  The amplitude of the spikes appears constant, before sharply decreasing until completely disappearing. Once the neuron stops spiking, it does not start spiking again if the input does not change. 
Our measurements indicate that, if pulses of the right current values are applied successively, the neuron will start spiking each time before eventually stopping. However, the duration of phasic bursting is  not always  the same even if the input is identical. \\

In order to quantify the effect, a statistical study of phasic bursting as a function of input current is presented in figure~\ref{Fig3}b. 
A current pulse is applied to the neuron, its output is recorded on the oscilloscope, and the average frequency during the pulse duration is then computed for each point. When the phenomenon of phasic bursting occurs, spikes stop during a fraction of the total duration of the pulse, which decreases the average frequency. Despite the apparent stochastic behavior, a clear trend in the  mean frequency evolution as a function of input emerges. For low currents, there is at first almost no phasic bursting, and the median frequency is almost equal to the maximum frequencies observed. Then as the input current increases, the proportion of phasic events increases and the median frequency decreases until no phasic bursting occurs.  \\

We  now present a theoretical analysis to determine the origin of the experimentally-observed phasic bursting.
We model our system with the circuit of figure~\ref{Fig1}b, neglecting the parasitic inductance and the intrinsic capacitance, that do not impact the qualitative neuron dynamics, in order to gain in simplicity and generality. The system is then simplified to two coupled first-order differential equations that link the voltage $V_d$ across the device and the temperature $T_d$ inside the active volume of the device. The first equation reads

\begin{equation}
    \frac{d V_d}{dt} = \frac{I_s}{C_{ext}} - \frac{V_d}{R_d C_{ext}}, \\
\label{eq: Kirchhoff}
\end{equation}
where 
$I_s$ is the input current and $R_d$ is the Poole-Frenkel resistance defined in equation~\ref{eq: R_d}.
The second equation is the Newton Law of Cooling (equation~\ref{Eq: Newton's Cooling Law}).

\begin{figure} [h]
\centering
\includegraphics[width=1\textwidth]{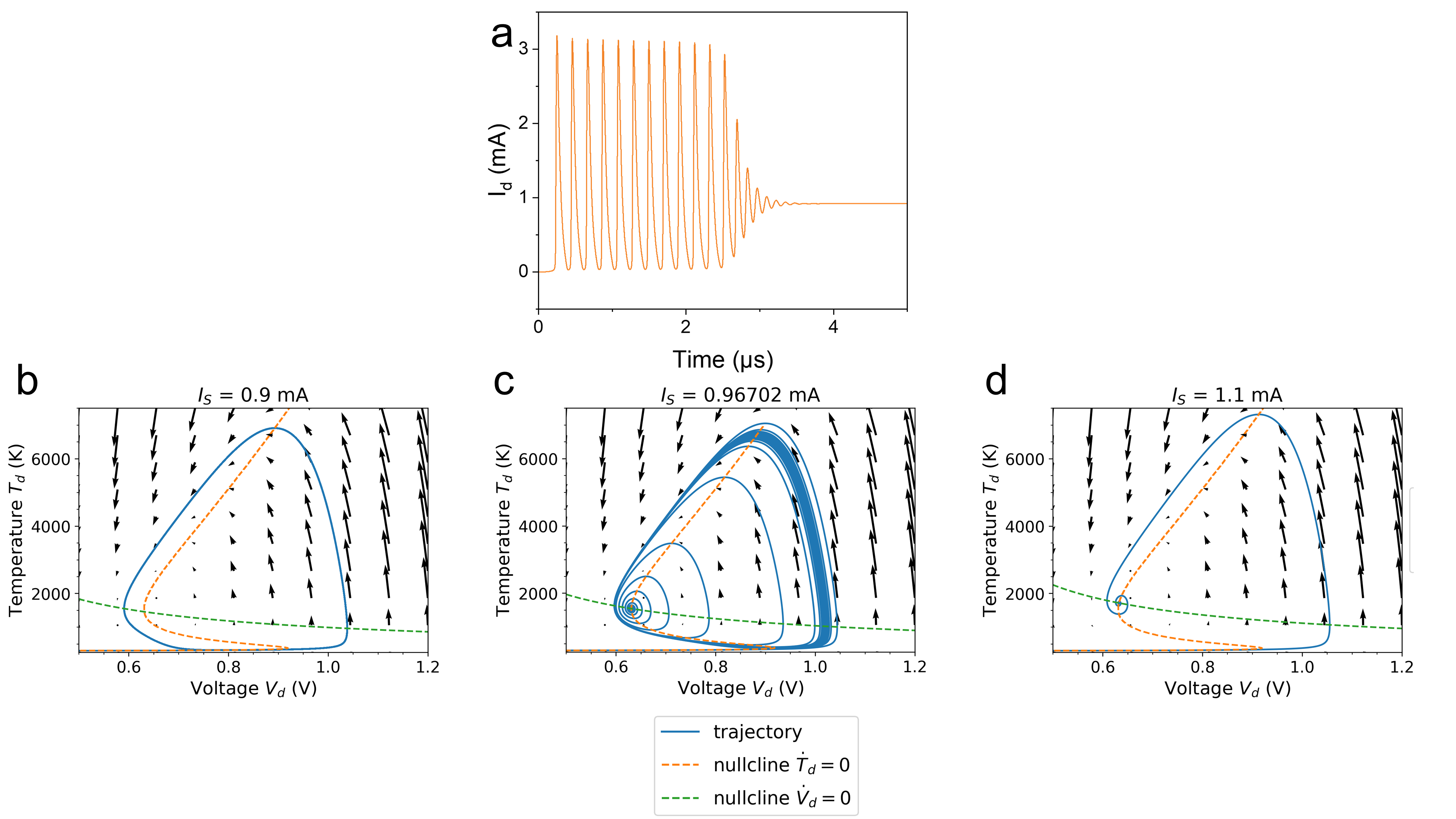}
\caption{\label{Fig4} a: Simulations of the device current oscillations as a function of
time for a current input $I_s$ of 0.96702 mA . b, c, d: simulation of the trajectory (in blue) and the nullclines (in orange for $\dot T =0$ and in green for $\dot V = 0$) for different input currents $I_s$ of value 0.9, 0.96702 and 1.1 mA for each figure. The y-axis corresponds to the temperature $T_d$ in the active volume of the device while the x-axis represents the voltage of the device $V_d$. The black arrows indicate the direction of the gradient at each point. }
\end{figure}

Equations \ref{Eq: Newton's Cooling Law} and \ref{eq: Kirchhoff}  can be solved numerically, leading to the different trajectories plotted in blue in figures~\ref{Fig4}b,c,d for the input current values $I_s$ of 0.9, 0.96702 and 1.A mA respectively. The system nullclines are also shown in dotted lines. These curves correspond to the zero values of the right-hand side of equations~\ref{eq: Kirchhoff} and~\ref{Eq: Newton's Cooling Law}. Their intersection in the two-dimensional phase space ($T_d$,$V_d$) corresponds to points for which the derivatives of $T_d$ and $V_d$ are zero, and therefore gives the fixed point of the system for each input current. 

Consistent with equation~\ref{Eq: Newton's Cooling Law},  the temperature nullcline does not depend on the input current $I_S$ and is therefore identical in figures~\ref{Fig4}b,c,d (orange curve).
On the other hand, increasing the input current vertically shifts the voltage nullcline to the top of the phase space. 
The current-dependent fixed points can therefore be obtained by following the temperature nullcline. 
For each of these points the Poole-Frenkel resistance can be computed, and by plotting the input current $I_s$ as the function of the voltage $V_d$ (thanks to the equilibrium relation $I_s = \frac{V_d}{R_d}$) the simulated quasistatic curve of figure~\ref{Fig1}a is obtained.\\

The analysis of figure 4 shows that phasic bursting is a particular situation that occurs around the hold point. Below the hold point, the fixed point is not stable, and the trajectory therefore reaches a limit cycle: this is what happens  in figure~\ref{Fig4}b. At the hold point, the system undergoes a supercritical Hopf bifurcation, where the limit cycle becomes a stable equilibrium point (as seen in figure~\ref{Fig4}d). Just above the transition (figure~\ref{Fig4}c), the system reaches a stable equilibrium point, but the convergence of the trajectory is quite slow (see figure~\ref{Fig4}a). This dynamic  naturally gives rise to the phasic bursting phenomenon of figure~\ref{Fig4}a, where an apparently stable train of spike unexpectedly fades out then stops.
Interestingly, in the experiments, the current input range where the phasic bursting happens ($\Delta I$ = 0.04 mA) is about ten times larger than in the simulations ($\Delta I$ = 0.003 mA). The noise inherent to physical devices and to the input current (close to 0.018 mA in our experiments)  explains the experimentally observed stochasticity of phasic bursting and expands the phasic bursting range. Indeed, even if the bias conditions of the device are set outside of the narrow range where phasic bursting is predicted in the absence of noise, fluctuations will enable the system to reach it and initiate the bifurcation, a phenomenon akin to stochastic resonance observed in biological neurons \cite{mcdonnell2009stochastic}. Other factors can also impact the details of the phasic bursting behavior. In the model, the thermal resistance is considered constant for simplicity, but this is not true in a real device. 

\section{Conclusion}
Volatile $NbO_x$ memristors are excellent neuron candidates as  they are scalable,  present reliable threshold switching, and are compatible with memristive synapses such as $HfO_2$ Metal-Insulator-Metal structures. We have shown that the $Pt/Nb_2O_5/Ti/Pt$ stack presents  well-suited I-V characteristics: a current-controlled S-shaped Negative Differential Resistance, which can be modeled by  assuming Poole-Frenkel conduction. This type of devices is able to spike and the resulting  shape is very close  to the one of a biological neuron with initial depolarization followed by hyperpolarization due to an inductance. We demonstrated that this device presents multiple computational properties such as Leaky-Integrate-and-Fire (LIF) characteristics, all-or-nothing-firing, and phasic bursting. We also investigated the origin of phasic bursting through the analysis of the physical equations of the devices. 
This phenomenon comes from the bifurcation between an unstable fixed point (limit cycle) and a stable fixed point (equilibrium) driven by Poole-Frenkel dynamics.
These results pave the way to easily-scalable neurons that can be easily modelled and simulated but still show a complex behavior in order to mimic biological computations.

\section{Acknowledgements}
This work is supported by the European Research Council Starting Grant NANOINFER: 715872 and by the Agence Nationale de la Recherche project ANR-20-CE24-0002 SpinSpike). We would also like to acknowledge access to NCRIS facilities at the ACT node of the Australian National Fabrication Facility (ANFF) and the Australian Facility for Advanced ion-implantation Research (AFAiiR), a node of the Heavy-Ion Accelerator Capability.

\newpage
\newpage
\newpage
\newpage

\bibliographystyle{unsrt}
\bibliography{biblio2.bib}
\newpage

\section*{Supplementary Material: Characterization and modeling of spiking and bursting in experimental $\textrm{NbO}_\textrm{x}$ neuron }

\begin{figure} [h]
\centering
\includegraphics[width=0.5\textwidth]{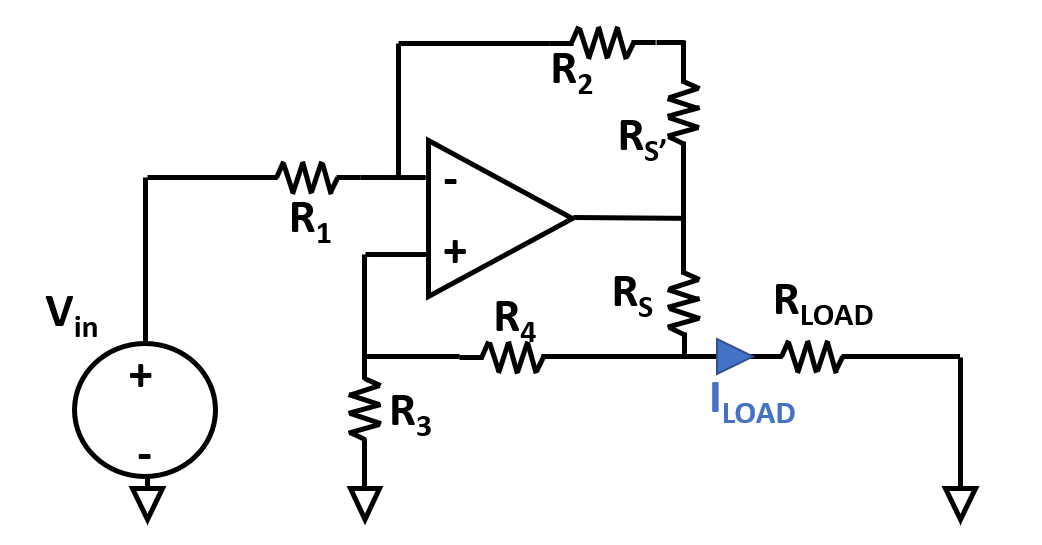}
\caption{Schematics of the voltage pulse to current pulse converter used in the experiments. Here,  $R_1 = R_3 = 1~k \Omega$, $R_2 = R_4 = 100~\Omega$, $R_S = R_{S'} = 400 \Omega$.}
\label{voltage-current_converter}
\end{figure}

\begin{table}[h]
\begin{center}
\begin{tabular}{ |c|c|c| } 
\hline
Variable & Value \\
\hline
$C_{ext}$ & 200 pF \\ 
$L_{ext}$ & 700 nH  \\ 
$R_{out}$ & 25 $\Omega$  \\ 
$C_{d}$ & 0.33 pF  \\ 
$R_{0}$ & 190 $\Omega$ \\ 
$E_{a}$ & 0.215 eV  \\ 
$\epsilon_r$ & 45  \\ 
d & 31 nm  \\ 
$C_{th}$ & 2e-15 $J \cdot K^{-1}$  \\ 
$R_{th}$ & 2040816   $K \cdot W^{-1}$\\ 
\hline
\end{tabular}

\caption{Table of the parameters used in both LTSpice and Python simulations.}
\end{center}
\label{table:1} 
\end{table}
\hypertarget{table}{~}

\begin{figure} [h]
\centering
\includegraphics[width=1\textwidth]{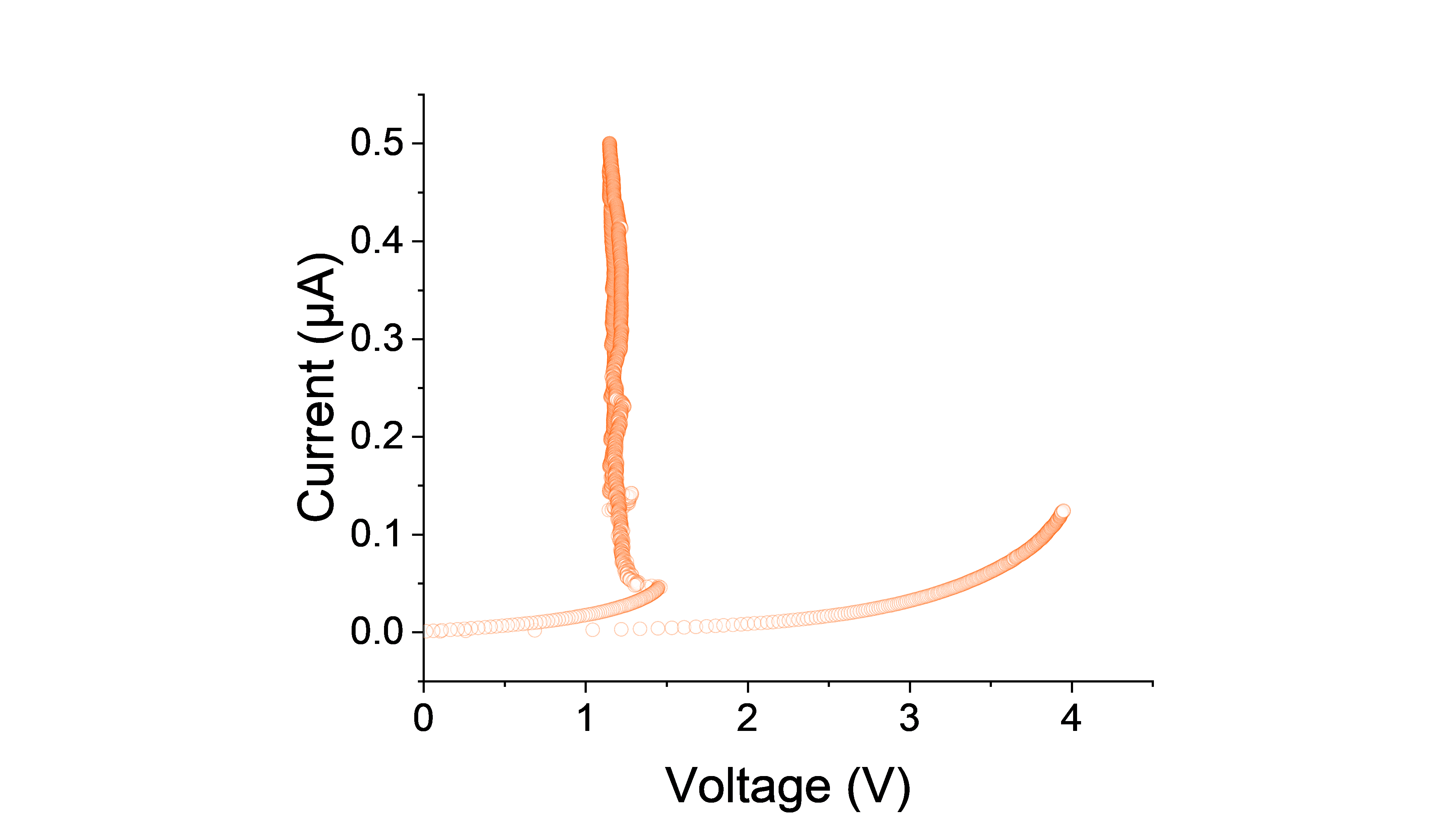}
\caption{Positive current-controlled electroforming with input current going from 0 to 0.5 mA.}
\label{electroforming}
\end{figure}

\begin{figure} [h]
\centering
\includegraphics[width=0.8\textwidth]{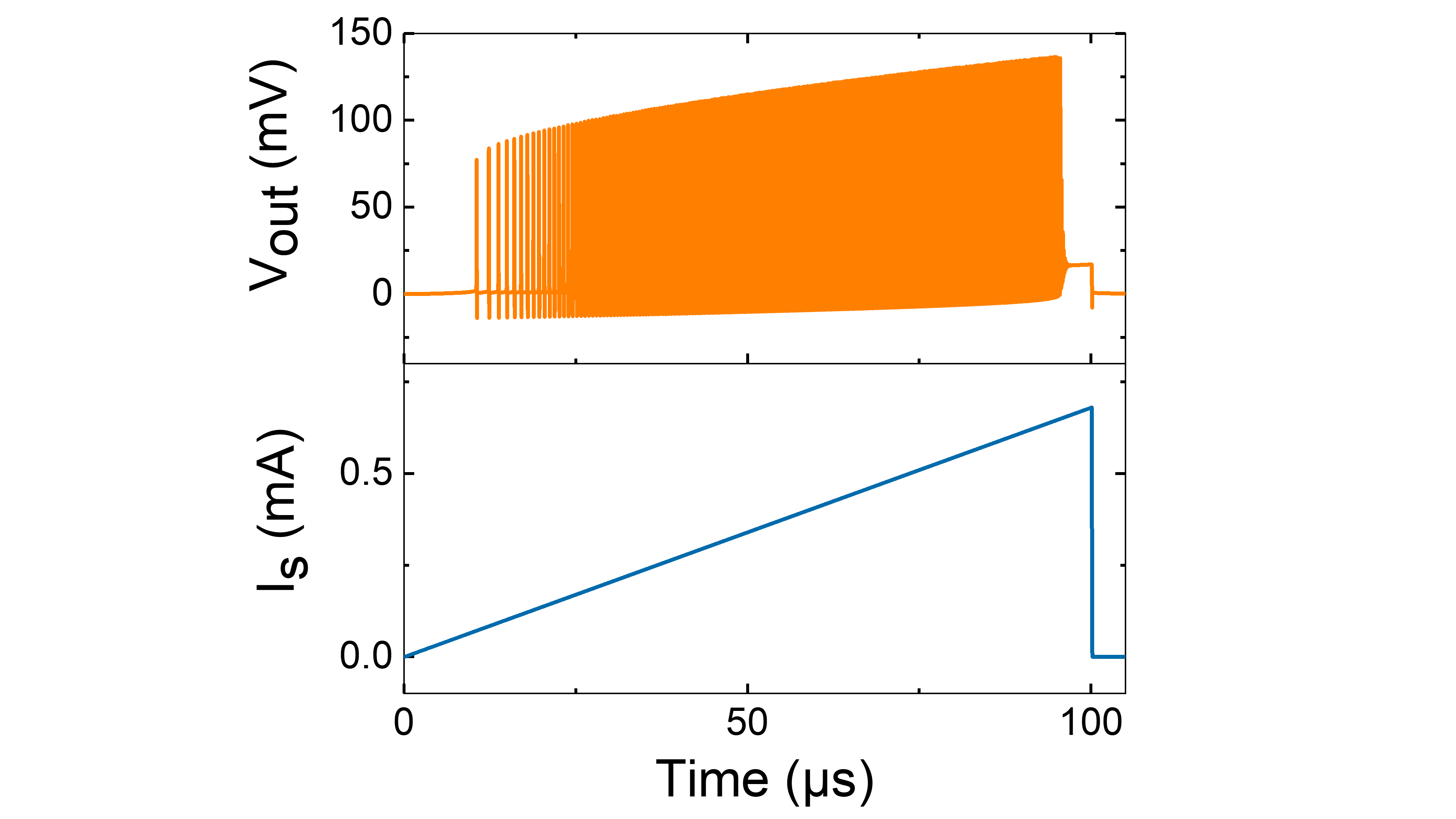}
\caption{Simulation of $NbO_x$ neuron output as a function of input current amplitude. A 100 $\mu$s current ramp from 0 to 680 $\mu$A and 100 ns  fall time is applied to the device. This simulation is realized in LTSpice, using the circuit shown in figure~\ref{Fig1}b and the parameters of table~\protect\hyperlink{table}{1}.}
\label{ramp}
\end{figure}

\begin{figure} [h]
\centering
\includegraphics[width=0.8\textwidth]{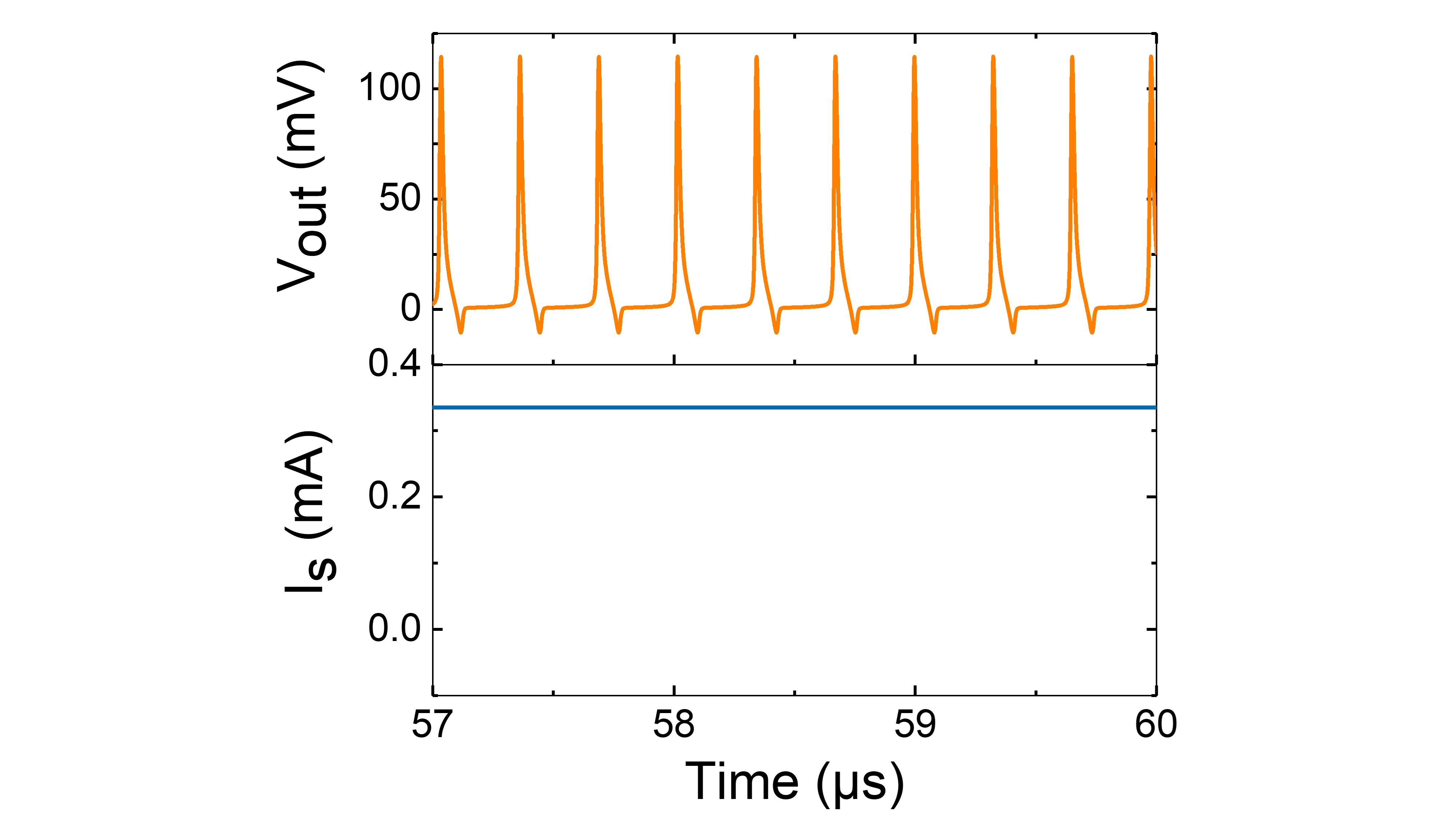}
\caption{Simulation of tonic spiking. The neuron receives an constant input current of 335 $\mu$A. This simulation is realized in LTSpice, using the circuit shown in figure~\ref{Fig1}b and the parameters of table~\protect\hyperlink{table}{1}.}
\label{tonic}
\end{figure}

\begin{figure} [h]
\centering
\includegraphics[width=0.8\textwidth]{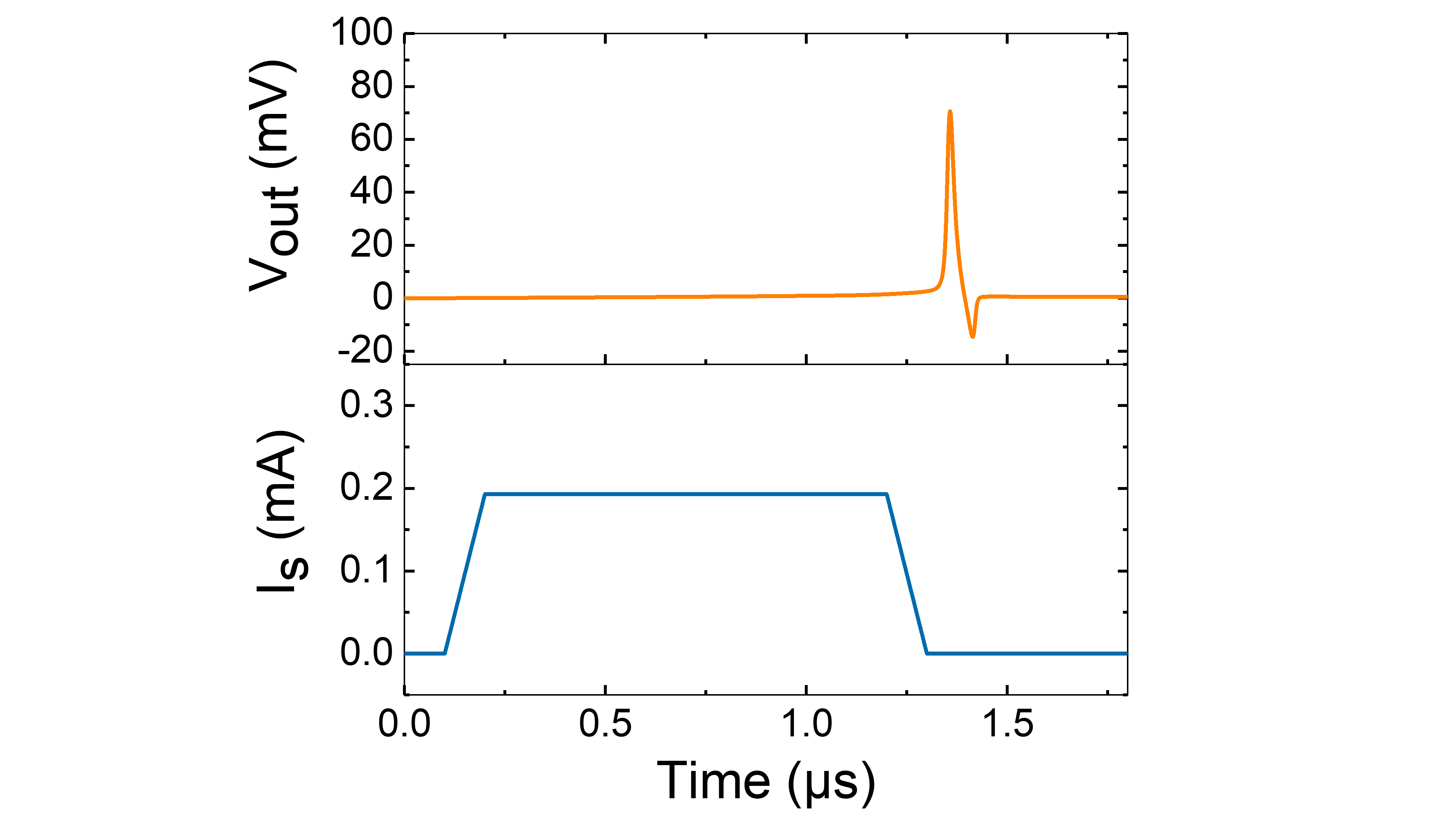}
\caption{Spike latency. A pulse of duration of 1
$\mu$s and value 193 $\mu$A with a rise time and fall time of both 100 ns is applied to
the neuron. This simulation is realized in LTSpice, using the circuit shown in figure~\ref{Fig1}b and the parameters of table~\protect\hyperlink{table}{1}.}
\label{latency}
\end{figure}

\begin{figure} [h]
\centering
\includegraphics[width=1\textwidth]{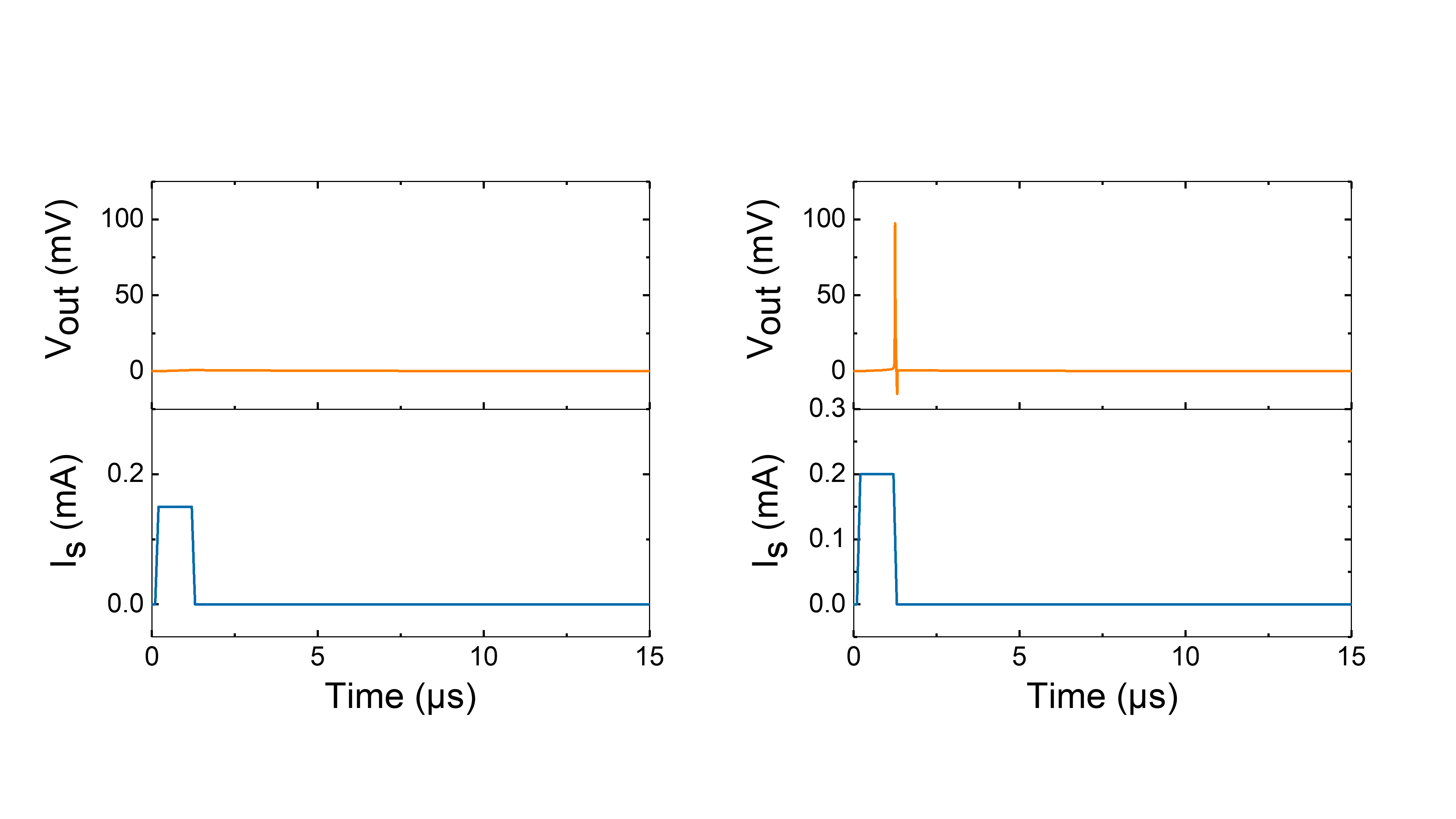}
\caption{Spatial integration. Comparison between two figures where a pulse of
duration of 1 $\mu$s with a rise time and fall time of both 100 ns are applied to the neuron.
On the left, the value of the current is 150 $\mu$A. On the right, the input current value
is 200 $\mu$A. These simulations are realized in LTSpice, using the circuit shown in figure~\ref{Fig1}b and the parameters of table~\protect\hyperlink{table}{1}.}
\label{all-or-nothing}
\end{figure}

\begin{figure} [h]
\centering
\includegraphics[width=1\textwidth]{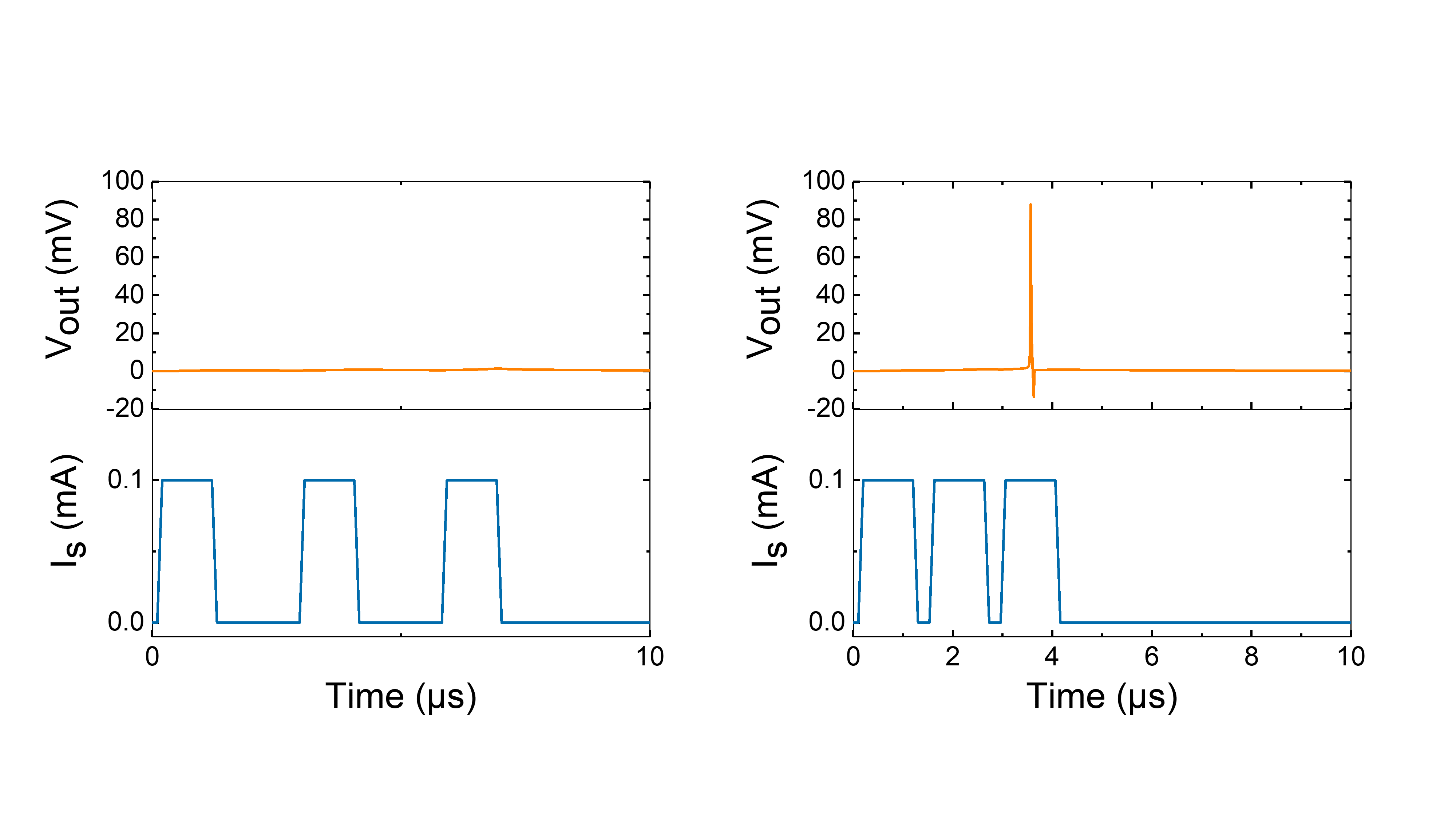}
\caption{Temporal integration. Three pulses of of duration of 1 $\mu$s with a rise
time and fall time of both 100 ns and of value 100 $\mu$A are applied to the neuron.
On the left, the time period is 2.86 $\mu$s (frequency of about 0.35 MHz). On the right, the time period is 1.43 $\mu$s (frequency of about 0.7 MHz). These simulations are realized in LTSpice, using the circuit shown in figure~\ref{Fig1} b and the parameters of table~\protect\hyperlink{table}{1}.}
\label{LIF}
\end{figure}

\end{document}